\documentclass[%
 reprint,
superscriptaddress,
 amsmath,amssymb,
 aps,
prl,
]{revtex4-2}
\UseRawInputEncoding
\usepackage{siunitx}
\DeclareSIUnit\atomicmassunit{u}
\usepackage{graphicx}
\usepackage{dcolumn}
\usepackage{bm}
\begin{document}

\preprint{APS/123-QED}

\title{The nuclear charge radius of $^{26m}$Al and its implication for V$_{ud}$ in the CKM matrix}

\author{P. Plattner}
    \email{Corresponding author: peter.plattner@cern.ch}
    \affiliation{ISOLDE, CERN Experimental Physics Department, Geneva 23, 1211 Genev\`e, Switzerland}
    \affiliation{Universit\"at Innsbruck, Innrain 52, 6020 Innsbruck, Austria}
    \affiliation{Max-Planck-Institut f\"ur Kernphysik, Saupfercheckweg 1, 69117 Heidelberg, Germany}
\author{E. Wood}
    \affiliation{Department of Physics, University of Liverpool, Liverpool, L69 7ZE, United Kingdom}
\author{L. Al Ayoubi}
    \affiliation{Department of Physics, University of Jyv\"askyl\"a, P.O. Box 35 FI-40014, Jyv\"askyl\"a, Finland}
\author{O. Beliuskina}
    \affiliation{Department of Physics, University of Jyv\"askyl\"a, P.O. Box 35 FI-40014, Jyv\"askyl\"a, Finland}
\author{M.L. Bissell}
    \affiliation{Department of Physics and Astronomy, University of Manchester, Oxford Road, Manchester, M13 9PL, United Kingdom}
    \affiliation{ISOLDE, CERN Experimental Physics Department, Geneva 23, 1211 Genev\`e, Switzerland}
\author{K. Blaum}
    \affiliation{Max-Planck-Institut f\"ur Kernphysik, Saupfercheckweg 1, 69117 Heidelberg, Germany}
\author{P. Campbell}
    \affiliation{Department of Physics and Astronomy, University of Manchester, Oxford Road, Manchester, M13 9PL, United Kingdom}
\author{B. Cheal}
    \affiliation{Department of Physics, University of Liverpool, Liverpool, L69 7ZE, United Kingdom}
\author{R.P. de Groote}
    \altaffiliation{Present address: Instituut voor Kern- en Stralingsfysica, KU Leuven, 3001 Leuven, Belgium}
    \affiliation{Department of Physics, University of Jyv\"askyl\"a, P.O. Box 35 FI-40014, Jyv\"askyl\"a, Finland}
\author{C.S. Devlin}
    \affiliation{Department of Physics, University of Liverpool, Liverpool, L69 7ZE, United Kingdom}
\author{T. Eronen}
    \affiliation{Department of Physics, University of Jyv\"askyl\"a, P.O. Box 35 FI-40014, Jyv\"askyl\"a, Finland}
\author{L. Filippin}
    \affiliation{Spectroscopy, Quantum Chemistry and Atmospheric Remote Sensing (SQUARES), Universit\'e libre de Bruxelles, 1050 Brussels, Belgium}
\author{R.F. Garc\'ia Ru\'iz}
    \affiliation{ISOLDE, CERN Experimental Physics Department, Geneva 23, 1211 Genev\`e, Switzerland}
    \affiliation{Massachusetts Institute of Technology, 77 Massachusetts Ave, Cambridge, MA 02139, USA}
\author{Z. Ge}
    \affiliation{Department of Physics, University of Jyv\"askyl\"a, P.O. Box 35 FI-40014, Jyv\"askyl\"a, Finland}
\author{S. Geldhof}
    \affiliation{Instituut voor Kern- en Stralingsfysica, KU Leuven, 3001 Leuven, Belgium}
\author{W. Gins}
    \affiliation{Department of Physics, University of Jyv\"askyl\"a, P.O. Box 35 FI-40014, Jyv\"askyl\"a, Finland}
\author{M. Godefroid}
    \affiliation{Spectroscopy, Quantum Chemistry and Atmospheric Remote Sensing (SQUARES), Universit\'e libre de Bruxelles, 1050 Brussels, Belgium}
\author{H. Heylen}
    \affiliation{ISOLDE, CERN Experimental Physics Department, Geneva 23, 1211 Genev\`e, Switzerland}
    \affiliation{Max-Planck-Institut f\"ur Kernphysik, Saupfercheckweg 1, 69117 Heidelberg, Germany}
\author{M. Hukkanen}
    \affiliation{Department of Physics, University of Jyv\"askyl\"a, P.O. Box 35 FI-40014, Jyv\"askyl\"a, Finland}
\author{P. Imgram}
    \affiliation{Institut f\"ur Kernphysik, Technische Universit\"at Darmstadt, Schlossgartenstra\ss e 9, 64289 Darmstadt, Germany}
\author{A. Jaries}
    \affiliation{Department of Physics, University of Jyv\"askyl\"a, P.O. Box 35 FI-40014, Jyv\"askyl\"a, Finland}
\author{A. Jokinen}
    \affiliation{Department of Physics, University of Jyv\"askyl\"a, P.O. Box 35 FI-40014, Jyv\"askyl\"a, Finland}
\author{A. Kanellakopoulos}
    \affiliation{Instituut voor Kern- en Stralingsfysica, KU Leuven, 3001 Leuven, Belgium}
\author{A. Kankainen}
    \affiliation{Department of Physics, University of Jyv\"askyl\"a, P.O. Box 35 FI-40014, Jyv\"askyl\"a, Finland}
\author{S. Kaufmann}
    \affiliation{Institut f\"ur Kernphysik, Technische Universit\"at Darmstadt, Schlossgartenstra\ss e 9, 64289 Darmstadt, Germany}
\author{K. K\"onig}
    \affiliation{Institut f\"ur Kernphysik, Technische Universit\"at Darmstadt, Schlossgartenstra\ss e 9, 64289 Darmstadt, Germany}
\author{\'A. Koszor\'us}
    \altaffiliation{Present address: Instituut voor Kern- en Stralingsfysica, KU Leuven, 3001 Leuven, Belgium}
    \affiliation{Department of Physics, University of Liverpool, Liverpool, L69 7ZE, United Kingdom}
    \affiliation{Instituut voor Kern- en Stralingsfysica, KU Leuven, 3001 Leuven, Belgium}
\author{S. Kujanp\"a\"a}
    \affiliation{Department of Physics, University of Jyv\"askyl\"a, P.O. Box 35 FI-40014, Jyv\"askyl\"a, Finland}
\author{S. Lechner}
    \affiliation{ISOLDE, CERN Experimental Physics Department, Geneva 23, 1211 Genev\`e, Switzerland}
\author{S. Malbrunot-Ettenauer}
    \email{Corresponding author: stephan.ettenauer@cern.ch}
    \affiliation{ISOLDE, CERN Experimental Physics Department, Geneva 23, 1211 Genev\`e, Switzerland}
    \affiliation{TRIUMF, 4004 Wesbrook Mall, Vancouver, BC V6T 2A3, Canada}
\author{P. M\"uller}
    \affiliation{Institut f\"ur Kernphysik, Technische Universit\"at Darmstadt, Schlossgartenstra\ss e 9, 64289 Darmstadt, Germany}
\author{R. Mathieson}
    \affiliation{Department of Physics, University of Liverpool, Liverpool, L69 7ZE, United Kingdom}
\author{I. Moore}
    \affiliation{Department of Physics, University of Jyv\"askyl\"a, P.O. Box 35 FI-40014, Jyv\"askyl\"a, Finland}
\author{W. N\"ortersh\"auser}
    \affiliation{Institut f\"ur Kernphysik, Technische Universit\"at Darmstadt, Schlossgartenstra\ss e 9, 64289 Darmstadt, Germany}
\author{D. Nesterenko}
    \affiliation{Department of Physics, University of Jyv\"askyl\"a, P.O. Box 35 FI-40014, Jyv\"askyl\"a, Finland}
\author{R. Neugart}
    \affiliation{Max-Planck-Institut f\"ur Kernphysik, Saupfercheckweg 1, 69117 Heidelberg, Germany}
    \affiliation{Institut f\"ur Kernchemie, Universit\"at Mainz, Fritz-Stra\ss mann-Weg 2, 55128 Mainz, Germany}
\author{G. Neyens}
    \affiliation{ISOLDE, CERN Experimental Physics Department, Geneva 23, 1211 Genev\`e, Switzerland}
    \affiliation{Instituut voor Kern- en Stralingsfysica, KU Leuven, 3001 Leuven, Belgium}
\author{A. Ortiz-Cortes}
    \affiliation{Department of Physics, University of Jyv\"askyl\"a, P.O. Box 35 FI-40014, Jyv\"askyl\"a, Finland}
\author{H. Penttil\"a}
    \affiliation{Department of Physics, University of Jyv\"askyl\"a, P.O. Box 35 FI-40014, Jyv\"askyl\"a, Finland}
\author{I. Pohjalainen}
    \affiliation{Department of Physics, University of Jyv\"askyl\"a, P.O. Box 35 FI-40014, Jyv\"askyl\"a, Finland}
\author{A. Raggio}
    \affiliation{Department of Physics, University of Jyv\"askyl\"a, P.O. Box 35 FI-40014, Jyv\"askyl\"a, Finland}
\author{M. Reponen}
    \affiliation{Department of Physics, University of Jyv\"askyl\"a, P.O. Box 35 FI-40014, Jyv\"askyl\"a, Finland}
\author{S. Rinta-Antila}
    \affiliation{Department of Physics, University of Jyv\"askyl\"a, P.O. Box 35 FI-40014, Jyv\"askyl\"a, Finland}
\author{L.V. Rodr\'iguez}
    \affiliation{Max-Planck-Institut f\"ur Kernphysik, Saupfercheckweg 1, 69117 Heidelberg, Germany}
    \affiliation{ISOLDE, CERN Experimental Physics Department, Geneva 23, 1211 Genev\`e, Switzerland}
    \affiliation{IJCLab, CNRS/IN2P3, Universit\'e Paris-Saclay, 91400 Orsay, France}
\author{J. Romero}
    \affiliation{Department of Physics, University of Jyv\"askyl\"a, P.O. Box 35 FI-40014, Jyv\"askyl\"a, Finland}
\author{R. S\'anchez}
    \affiliation{GSI Helmholtzzentrum f\"ur Schwerionenforschung, Planckstra\ss e 1, 64291 Darmstadt, Germany }
\author{F. Sommer}
    \affiliation{Institut f\"ur Kernphysik, Technische Universit\"at Darmstadt, Schlossgartenstra\ss e 9, 64289 Darmstadt, Germany}
\author{M. Stryjczyk}
    \affiliation{Department of Physics, University of Jyv\"askyl\"a, P.O. Box 35 FI-40014, Jyv\"askyl\"a, Finland}
\author{V. Virtanen}
    \affiliation{Department of Physics, University of Jyv\"askyl\"a, P.O. Box 35 FI-40014, Jyv\"askyl\"a, Finland}
\author{L. Xie}
    \affiliation{Department of Physics and Astronomy, University of Manchester, Oxford Road, Manchester, M13 9PL, United Kingdom}
\author{Z.Y. Xu}
    \affiliation{Instituut voor Kern- en Stralingsfysica, KU Leuven, 3001 Leuven, Belgium}
\author{X. F. Yang}
    \affiliation{Instituut voor Kern- en Stralingsfysica, KU Leuven, 3001 Leuven, Belgium}
    \affiliation{School of Physics and State Key Laboratory of Nuclear Physics and Technology, Peking University, 209 Chengfu Road, 100871 Beijing, China}
\author{D.T. Yordanov}
    \affiliation{IJCLab, CNRS/IN2P3, Universit\'e Paris-Saclay, 91400 Orsay, France}

\date{\today}

\begin{abstract}
Collinear laser spectroscopy was performed on the isomer of the aluminium isotope $^{26m}$Al. The measured isotope shift to $^{27}$Al in the $3s^{2}3p\;^{2}\!P^\circ_{3/2} \rightarrow 3s^{2}4s\;^{2}\!S_{1/2}$  atomic transition enabled the first experimental determination of the nuclear charge radius of $^{26m}$Al, resulting in $R_c$=\qty{3.130\pm.015}{\femto\meter}. This differs by 4.5 standard deviations from the extrapolated value used to calculate the isospin-symmetry breaking corrections in the superallowed $\beta$ decay of $^{26m}$Al. Its corrected $\mathcal{F}t$ value, important for the estimation of $V_{ud}$ in the CKM matrix, is thus shifted by one standard deviation to \qty{3071.4\pm1.0}{\second}.

\end{abstract}

\maketitle

{\it Introduction. ---}
The Cabibbo-Kobayashi-Maskawa (CKM) matrix is a central cornerstone in the formulation of the Standard Model of particle physics. It connects the quarks' mass with weak eigenstates and, thus, characterises the strength of quark-flavour mixing through the weak interaction. The first element in the top row of the matrix, $V_{ud}$, manifests in the $\beta$ decay of pions, neutrons or radioactive nuclei.
While individual entries of the quark mixing matrix cannot be predicted within the Standard Model, the CKM matrix is required to be unitary -- a tenet which is the subject of intense experimental scrutiny.

In recent years, the unitary test of the top-row elements: 
$$|V_{ud}|^2+|V_{us}|^2+|V_{ub}|^2=1-\Delta_{CKM}$$
has received significant attention. The unitarity of the CKM matrix demands the residual $\Delta_{CKM}$ to vanish. However, recent advances in the theoretical description of (inner) radiative corrections \cite{PhysRevLett.121.241804,PhysRevD.100.013001,PhysRevD.100.073008,PhysRevD.101.111301,PhysRevD.103.113001,PhysRevD.104.033003} to $\beta$ decays resulted in a notable shift in $V_{ud}$ and, thus, to a tension with respect to CKM unitarity. Following recommended values by the Particle Data Group~\cite{PDG2022}, $\Delta_{CKM}=\num{15\pm7e-4}$ hints at a $\approx\qty{2}{\sigma}$ deviation from unitarity although this discrepancy could be as large as \qty{5.5}{\sigma}, depending on which calculation of (nuclear-structure dependent
\cite{TOWNER199413,PhysRevD.100.013001,PhysRevLett.123.042503,PhysRevC.102.045501} and universal \cite{PhysRevLett.121.241804,PhysRevD.100.013001,PhysRevD.100.073008,PhysRevD.104.033003}) radiative corrections are used in the determination of $V_{ud}$ and which
decay is considered to obtain $V_{us}$ \cite{doi:10.1146/annurev.nucl.53.013103.155258, PhysRevLett.92.251803, AMBROSINO200676, Antonelli2010,  Amhis2021, FlavourLatticeAveragingGroupFLAG:2021npn, PDG2022}.

At present, superallowed $0^+\rightarrow 0^+$ nuclear $\beta$ decays remain the most precise way to access $V_{ud}$~\cite{PhysRevC.102.045501}. For these cases, the experimentally measured $ft$ value, characterising a $\beta$~decay, can be related to a corrected $\mathcal{F}t$ value:
\begin{equation}\label{eq:corrFt}
\mathcal{F}t = ft\cdot(1+\delta^\prime_R)(1+\delta_{NS}-\delta_{C}),
\end{equation}
where $\delta^\prime_R$ and $\delta_{NS}$ constitute the transition-dependent contributions to the radiative corrections while $\delta_C$ are the isospin-symmetry breaking (ISB) corrections. According to the conserved vector-current hypotheses, the $\mathcal{F}t$ values should be identical for all superallowed $\beta$~decays. When averaged over all 15 precision cases, they serve to extract $V_{ud}$.

While the experimental dataset on $ft$ values of superallowed $\beta$ decays robustly builds on 222 individual measurements~\cite{PhysRevC.102.045501}, theoretical corrections are 
under scrutiny. As part of this process, the uncertainties in the nuclear-structure dependent radiative corrections $\delta_{NS}$ have recently been inflated by a factor of $\approx2.6$~\cite{PhysRevC.102.045501}. Moreover, the ISB corrections $\delta_C$, which are also nuclear-structure dependent, remain an ongoing focus of research which has stimulated new theoretical calculations \cite{PhysRevC.94.024306,PhysRevC.97.024324,PhysRevC.104.014324,SENG2023137654} as well as experimental benchmarks \cite{PhysRevLett.107.182301,PhysRevC.82.065501,PhysRevLett.112.102502,PhysRevC.91.045504,PhysRevC.100.015503,PhysRevC.101.045501}. 

For the determination of $V_{ud}$, $^{26m}$Al is of particular importance. The nuclear-structure dependent corrections, $\delta_{NS}-\delta_{C}$, in $^{26m}$Al are the smallest in size among all superallowed $\beta$ emitters \cite{PhysRevC.102.045501}. The same holds true for the combined experimental and theoretical uncertainties in the $\mathcal{F}t$ value of $^{26m}$Al\cite{PhysRevC.102.045501}. Its extraordinary precision is thus almost on par with all other precision cases combined. In times of tension with CKM unitary and rigorous examination of all involved theoretical corrections, it is, therefore, unsettling that one critical input parameter for the calculation of $\delta_C$, i.e. the nuclear charge radius, is in the case of $^{26m}$Al, based on an extrapolated but experimentally unknown value \cite{PhysRevC.66.035501,PhysRevC.77.025501}. 

In this Letter, we report on isotope-shift measurements obtained via collinear laser spectroscopy (CLS) that puts the nuclear charge radius of $^{26m}$Al on solid experimental footings. Implications for its $\mathcal{F}t$ value and, thus, $V_{ud}$ are discussed. 

{\it Experiment. ---}
Two independent experiments were performed, one at the COLLAPS 
beamline \cite{Neugart2017CollinearHighlights} at ISOLDE/CERN \cite{Catherall_2017} and the other at the IGISOL CLS beamline \cite{PhysRevA.97.042504} in Jyv\"askyl\"a/Finland.
Details of the campaign on aluminium isotopes at COLLAPS are described in Ref.~\cite{ Heylen2021}. In short, radioactive aluminium atoms were synthesised by bombarding a uranium carbide target with 1.4-GeV protons from CERN's PS booster. Once released from the production target, the Al$^+$ ion beam was formed via resonant laser ionisation \cite{Fedosseev2017IonISOLDE}, subsequent electrostatic acceleration to 30 keV, and final mass selection via ISOLDE's magnetic high-resolution separator \cite{GILES2003497}. 

At IGISOL \cite{Moore2014}, the radionuclides of interest were produced in $^{27}$Al(p,d) reactions at 25-MeV proton energy. After their release from a thin foil target and extraction from the He-gas filled gas cell, the Al ions were guided
towards the high vacuum region of the mass separator via a sextupole ion guide, accelerated to \qty{30}{\kilo\eV} and mass separated by a $55^{\circ}$ dipole magnet.

In both experiments, the ions were stopped, cooled and accumulated in a buffer-gas filled radio-frequency-quadrupole cooler buncher \cite{PhysRevLett.88.094801,FRANBERG20084502} before they were delivered in 

30-keV ion bunches to the respective CLS beamline.  
There, the ion beam was spatially super-imposed with the 
laser beam in collinear (COLLAPS) or anti-collinear (IGISOL) fashion. The ions' velocity was adjusted by a Doppler-tuning voltage applied before the neutralisation in a charge exchange cell filled with sodium vapour. In this manner, the laser frequency experienced in the rest frame of the neutral Al atoms could be scanned via Doppler tuning. Once on resonance with the selected transition, fluorescence was detected using a series of photomultiplier tubes and their associated lens systems which surrounded the laser-atom interaction region \cite{Kreim2014NuclearN=28, Groote2020}. 

In both campaigns, the main spectroscopic transition was from the atomic $3s^{2}3p\;^{2}\!P^\circ_{3/2} \rightarrow 3s^{2}4s\;^{2}\!S_{1/2}$ level at \qty{25235.695}{\per\centi\meter}. Suitable laser light was provided by a continuous wave Ti:Sa ring laser (Sirah Matisse 2) set to an output wavelength of \qty{792}{\nano\meter}. The resulting laser light was frequency doubled using an external cavity frequency doubler (Wavetrain 2), after which the laser beam with a few milliwatts in power was sent through the experimental beamlines. To compensate for long-term drifts in both experiments, the fundamental laser was locked in wavelength to a HighFinesse WSU-10 wavemeter which was regularly calibrated to a frequency-stabilised HeNe laser.

    \begin{figure*}
        \centering
        \includegraphics[width=\textwidth]{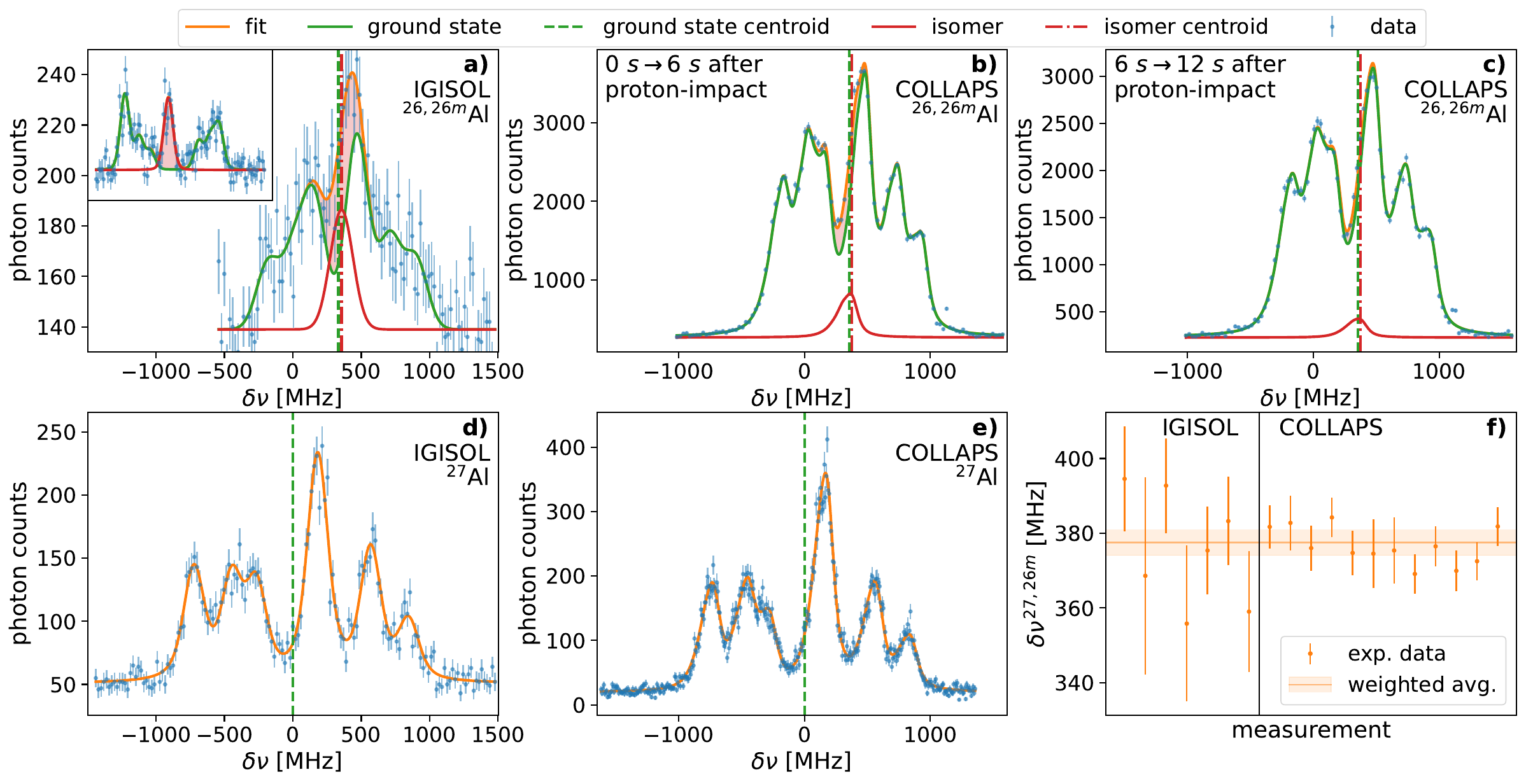}
        \caption{\textbf{(a)} Example of a resonance spectrum of the main spectroscopic transition $3s^{2}3p\;^{2}\!P^\circ_{3/2} \rightarrow 3s^{2}4s\;^{2}\!S_{1/2}$ obtained in the CLS measurements of $^{26,26m}$Al at IGISOL. The inset demonstrates the isomer's presence (red) due to well separated ground and isomer states in the $3s^{2}3p\;^{2}\!P^\circ_{1/2} \rightarrow 3s^{2}3d\;^{2}\!D_{3/2}$ transition. \textbf{(b,c)} The spectra of $^{26,26m}$Al in the main transition at COLLAPS. Ions have been extracted \qty{0}{\second} (b) and \qty{6}{\second} (c) after the proton impact on the ISOLDE target, demonstrating the isomer's presence due to the decrease in intensity consistent with the isomer's half-life. \textbf{(d,e)} Examples of resonance spectra of the $^{27}$Al references studied using the main transition at IGISOL (d) and COLLAPS (e). \textbf{f)} Extracted isotope shifts (points) and the resulting weighted average (a horizontal line), including systematic uncertainties.}
        \label{fig:example_spectra}
    \end{figure*}

In addition to the low-lying $^{26m}$Al isomer with a half-life of $T_{1/2}=\qty{6.34602+-.00054}{\second}$ \cite{PhysRevC.102.045501}, the long-lived ground state $^{26}$Al ($T_{1/2}=\qty{7.17+-.24e5}{y}$ \cite{26AlGSHL}) was also present in the radioactive ion beams, although in different relative intensities which reflected the distinct production methods at ISOLDE and IGISOL. Examples of the obtained resonance spectra of $^{26,26m}$Al are shown in Fig.~\ref{fig:example_spectra}. Due to the dense hyperfine structure of $^{26}$Al (nuclear spin $I=5$) and the small isomer shift, the single peak associated with $^{26m}$Al ($I=0$) could not be resolved from the strongest transition in $^{26}$Al. In order to unambiguously demonstrate the presence of the isomer, the transition $3s^{2}3p\;^{2}\!P^\circ_{1/2} \rightarrow 3s^{2}3d\;^{2}\!D_{3/2}$ at \qty{308}{\nano\meter} was additionally utilised during the campaign at IGISOL. As visible in the inset of Fig.~\ref{fig:example_spectra}a, by exploiting the latter transition the multiplets in the hyperfine spectrum of $^{26}$Al were well separated (green line) and offered unobstructed access to the resonance peak of $^{26m}$Al (red). However, due to state-mixing with a second close-lying atomic state ($\Delta E\approx\qty{0.17}{\milli\eV}$), this transition is inadequate for the determination of nuclear charge radii \cite{Levins1997The26Al}.

To confirm the presence of the isomer in the ISOLDE beam, we took advantage of the pulsed time-structure of the proton beam and the long release time of Al from the ISOLDE target. In a set of dedicated measurements, subsequent proton pulses were separated in time by at least \qty{12}{\second} corresponding to approximately two half-lives of $^{26m}$Al. The recorded fluorescence data was divided into two sets which were measured up to \qty{6}{\second} and between 6 and \qty{12}{\second} after the proton impact, see Fig.~\ref{fig:example_spectra}b and \ref{fig:example_spectra}c, respectively. The resonances' intensities for the long-lived ground state (green) changed only slightly between these two data sets, likely because of a small time dependence in the Al release.  The much stronger decrease in isomer intensity (red) between the first and second \qty{6}{\second} of data taking was consistent with the isomer's half-life when each is normalised to the corresponding ground-state intensity. 

Direct comparison of the spectra shows a higher overall rate and thus better statistics for the COLLAPS data set. 
This statement holds true for both measurements of $^{26,26m}$Al as well as stable $^{27}$Al, see Fig.~\ref{fig:example_spectra}d and \ref{fig:example_spectra}e, which were interleaved with online data as reference measurements. On the other hand, the data from IGISOL benefits from a more favorable isomer-to-ground state ratio, compare Fig.~\ref{fig:example_spectra}a and \ref{fig:example_spectra}b. The complementarity of the COLLAPS and IGISOL data sets in terms of high statistics versus better isomer-ground state ratio was further strengthened by their distinct control and evaluation of systematic uncertainties. Most importantly, the determination of the ion-acceleration voltage at COLLAPS was achieved by a high-precision voltage divider. At IGISOL, it was calibrated by  CLS measurements of stable magnesium  (Mg) ions with respect to their precisely known isotope shifts. 

{\it Analysis and Results. ---}
The measured resonance spectra  of $^{26,26m,27}$Al were fitted to the theoretical model of the hyperfine spectra using the SATLAS package~\cite{Gins2018AnalysisPackage}. To constrain the fit in the present work, the ratio of the hyperfine parameters $A(P_{3/2})/A(S_{1/2})$ was fixed to the precise value of 4.5701(14), obtained in previous work on Al isotopes at COLLAPS \cite{Heylen2021}. However, this constraint was not applied to the present $^{27}$Al part of the COLLAPS analysis as the analysed spectra were a subset of the measurements examined in Ref.~\cite{Heylen2021}. 

For $^{26,26m}$Al, a model of the $I=5$ ground state and one of the $I=0$ isomeric state were superimposed. 
Within each experimental campaign, all $^{26,26m}$Al resonance spectra were fitted simultaneously with  the same, shared hyperfine parameters as long as a parameter was not otherwise constrained, see above. Similarly, the isomer shift between ground and isomeric state in $^{26}$Al was implemented as a shared fit parameter across a campaign's entire data set. The isomer centroid $\nu_0^{26m}$ itself was freely varied for each individual spectrum. For the determination of $\nu_0^{26m}$, the Doppler-tuning voltage was converted into frequency based on the isomer's ionic mass. It was verified in fits of simulated spectra that this approach led to accurate results despite the peak overlap with the resonance spectrum of the ground state. 

Voigt profiles were chosen for the lineshapes of individual resonance peaks with no intensity constraints in the ground state. The Lorentzian and Gaussian widths were shared between ground state and isomer peaks within each individual spectrum but not shared overall. Due to inelastic collisions in the charge-exchange cell \cite{Bendali1986Na+-NaSpectroscopy, Klose2012TestsSpectroscopy, Kreim2014NuclearN=28}, four equidistant side peaks were considered in the analysis of the COLLAPS data \cite{Heylen2021}. The energy offset of these sidepeaks was determined empirically and the relative intensities were constrained by Poisson's law. Because of lower statistics, the IGISOL data were found to be insensitive to the inclusion of these sidepeaks, thus, they were not considered in the analysis. 

Each spectrum of $^{26,26m}$Al was measured in sequence with an independent $^{27}$Al reference measurement. 
The isotope shift $\delta\nu^{27,26m}=\nu^{26m}_0-\nu^{27}_0$ of each measurement pair was calculated from the frequency centroid $\nu^{26m}_0$ of $^{26m}$Al with respect to the frequency centroid $\nu^{27}_0$ of the closest $^{27}$Al reference measurement. The results of all individual $\delta\nu^{27,26m}$ determinations are shown in Fig.~\ref{fig:example_spectra}f. Weighted averages in $\delta\nu^{27,26m}$ are calculated separately for the COLLAPS and IGISOL data sets, see Tab.~\ref{tab:isotope_shifts}.   

Systematic uncertainties in CLS for measurements of isotope shifts are well understood  \cite{Mueller1983SpinsSpectroscopy, KramerJ.2010ConstructionCERN, Krieger2011CalibrationSpectroscopy, Lechner2021LaserTrap} and are dominated by the imperfect knowledge of the beam energy. The acceleration voltage from the cooler-buncher at IGISOL was calibrated by matching measured isotope shifts in the D1 and D2 lines for singly-charged ions of stable magnesium isotopes to their precisely known literature values in Ref. \cite{Batteiger2009}. The remaining uncertainty in beam energy was 1.8 eV. An additional \num{1e-4} relative uncertainty was assigned to the scanning voltage in the Doppler tuning. For the COLLAPS data, a \num{1.5e-4} relative uncertainty of the incoming ion beam energy was assigned following the specifications of the employed voltage divider (Ohmlabs KV-30A). This was combined with the uncertainties of the calibrated  JRL KV10 voltage divider used to measure the scanning voltage and of the employed voltmeters (Agilent 34661A).  

Since the systematic uncertainties at COLLAPS and IGISOL were fully independent, statistical and systematic uncertainties of each measurement campaign were first added in quadrature before the weighted average of both measurement results was calculated, see Tab.~\ref{tab:isotope_shifts}.  Our final value for the isotope shift between $^{26m}$Al and $^{27}$Al is $\delta\nu^{27,26m}$=\qty{377.5\pm3.4}{\mega\hertz}.    
    \begin{table}[b]
    \caption{\label{tab:isotope_shifts}
    Measured isotope shift $\delta\nu^{27,26m}$ between $^{27}$Al and $^{26m}$Al obtained at the IGISOL facility and at COLLAPS/ISOLDE. The weighted average of the two measurements and the resulting difference in mean square charge radius $\delta\langle r_c^2\rangle^{27,26m}$ is listed.}
    \begin{ruledtabular}
    \begin{tabular}{rlc}
     & $\delta\nu^{27,26m}$ [\unit{\mega\hertz}] & $\delta\langle r_c^2\rangle^{27,26m}$ [fm$^2$]\\
    \hline
    COLLAPS & 376.5\{17\}[37]\footnotemark[1] & \\
    IGISOL & 379.7\{55\}[22]\footnotemark[1] & \\
    weighted average & 377.5(34)\footnotemark[2] & 0.429(45)\textlangle 76\textrangle\footnotemark[2]\\
    \end{tabular}
    \end{ruledtabular}
    \footnotetext[1]{Statistical and systematic uncertainties given in curly and square brackets, respectively.}
    \footnotetext[2]{Combined statistical and systematic uncertainties in parentheses. Uncertainty from atomic physics calculations of mass and field shift from \cite{Heylen2021} in angle brackets.}
    \end{table}

With knowledge of the isotope shift $\delta\nu^{27,26m}$ the difference in mean square nuclear charge radii $\delta\langle r^2\rangle$ between the two isotopes could be calculated according to \cite{Martensson1990}: $$\delta\nu^{27,26m} = F\delta\langle r^2\rangle^{27,26m}+M\frac{m_{26m}-m_{27}}{m_{27}(m_{26m}+m_e)},$$ where $m_e$ is the electron mass \cite{Sturm2014} and $m_A$ are the nuclear masses obtained when 13 electrons are subtracted from the atomic masses \cite{Huang2021TheProcedures} and an excitation energy of \qty{228.305}{\kilo\eV} \cite{ALKEMADE1982383} is added for $^{26m}$Al. Precision atomic-physics calculations were performed in a multiconfiguration Dirac-Hartree-Fock framework
to evaluate the field and mass shift factors $F$ and $M$ of the investigated atomic transition \cite{PhysRevA.94.062508,Heylen2021}. Combining the adopted values of $F$=\qty[per-mode=symbol]{76.2(22)}{\mega\hertz\per\femto\meter\squared} and $M$=\qty{-243(4)}{\giga\hertz\atomicmassunit} with the isotope shift $\delta\nu^{27,26m}$ of the present work yields $\delta\langle r^2\rangle_{27,26m}=\qty{0.429(88)}{\femto\meter\squared}$, see Tab.~\ref{tab:isotope_shifts}. Finally, the root mean square (rms) nuclear charge radius of $^{26m}$Al can be derived: $$R_{c}(^{26m}\mathrm{Al})\equiv\langle r^2\rangle_{26m}^{1/2}=\sqrt{R_{c}(^{27}\mathrm{Al})^2+\delta\langle r^2\rangle^{27,26m}}.$$
Using the previously evaluated rms charge radius of $^{27}$Al, $R_{c}(^{27}\mathrm{Al})$=\qty{3.061(6)}{\femto\meter}~\cite{Heylen2021}, a value of $R_{c}(^{26m}\mathrm{Al})$=\qty{3.130(15)} {\femto\meter} is obtained, see Tab.~\ref{tab:summary_results}.
    
{\it Discussion. ---} 
Nuclear charge radii of superallowed $\beta$ emitters are essential input parameters for the calculation of the ISB corrections $\delta_C$ when a nuclear shell-model approach with Woods-Saxon radial wavefunctions is employed \cite{PhysRevC.66.035501,PhysRevC.77.025501}. Currently, these $\delta_C$ calculations are the only ones considered to be sufficiently reliable to evaluate $\mathcal{F}t$ values and thus $V_{ud}$ \cite{PhysRevC.102.045501}.
In the shell-model approach, the ISB corrections are separated into two components, 
$\delta_C=\delta_{C1}+\delta_{C2}$. The former is associated with the configuration mixing within the restricted shell model space while the latter, known as the radial overlap correction, is derived from a phenomenological Woods-Saxon potential and it depends on the nuclear charge radius $R_c$.

\begin{figure}
        \centering
        \includegraphics[width=\columnwidth]{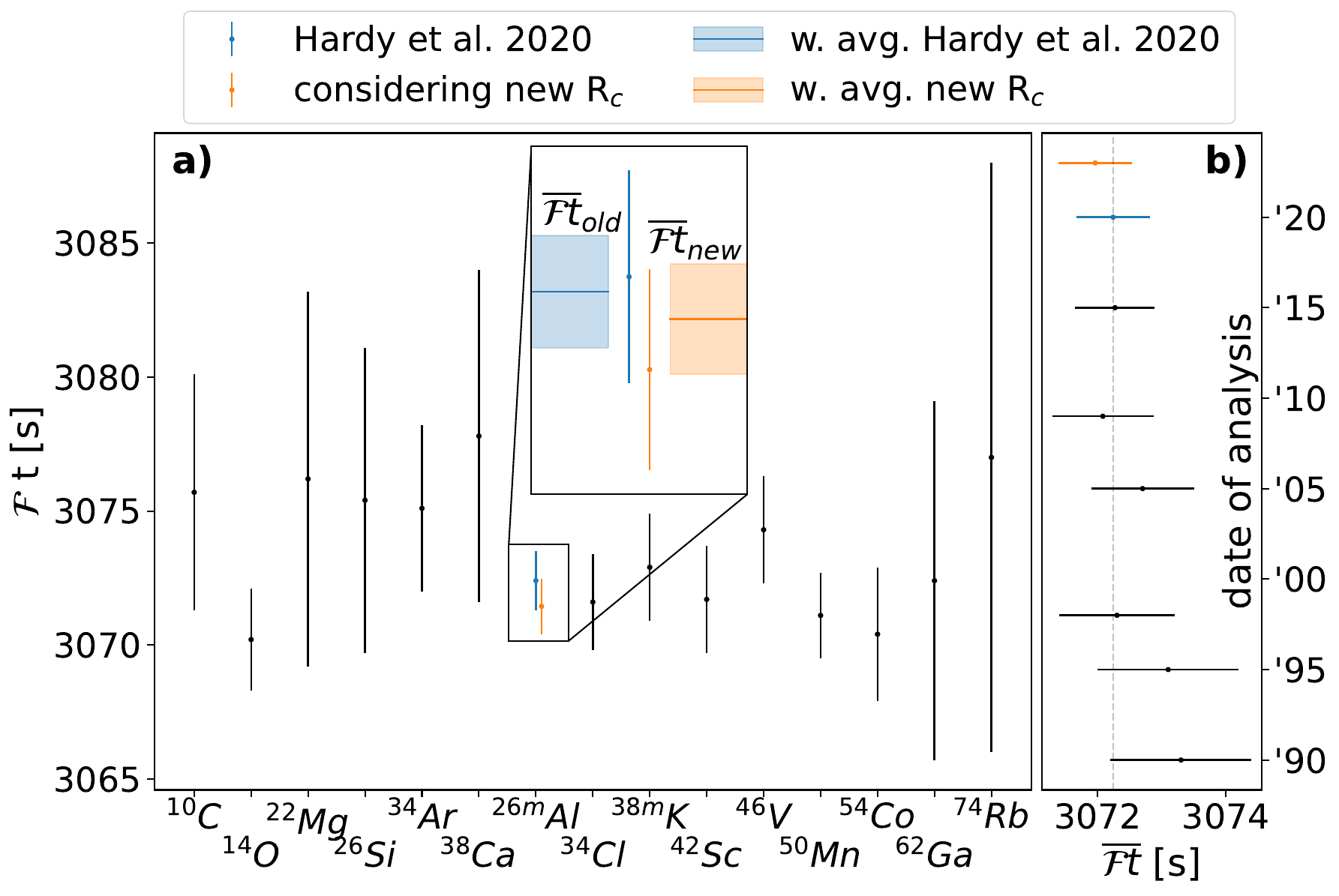}
        \caption{\textbf{(a)} $\mathcal{F}t$ values of the 15 superallowed $\beta$ emitters used to determine  V$_{ud}$. The values in black, taken from \cite{PhysRevC.102.045501}, include experimental as well as `statistical' theoretical errors. The previously determined  $\mathcal{F}t$ value for $^{26m}$Al \cite{PhysRevC.102.045501} (blue) is compared to the one (orange) when considering the experimental nuclear charge radius of the present work. The weighted averages for the 15 superallowed $\beta$ emitters are shown as horizontal bars in the inset (without considering additional, systematic theoretical uncertainties). \textbf{(b)} Evolution of the $\overline{\mathcal{F}t}$ value with statistical uncertainties in previous reviews \cite{HARDY1990429, Towner_1995, towner1998current, PhysRevLett.94.092502, PhysRevC.79.055502, PhysRevC.91.025501, PhysRevC.102.045501} (black) compared to this work (orange). The vertical line to guide the eye corresponds to the value from 2020 \cite{PhysRevC.102.045501}.}
        \label{fig:Ft_shift}
    \end{figure}

Since $R_c(^{26m}\mathrm{Al})$ was previously unknown, the calculation of $\delta_{C2}$ used $R_c$=\qty{3.040(20)}{\femto\meter} \cite{PhysRevC.66.035501}, an extrapolation based on other, known nuclear charge radii. Our experimental result, $R_c(^{26m}\mathrm{Al})$=\qty{3.130(15)}{\femto\meter}, deviates from this extrapolation by 4.5 standard deviations. This significantly impacts the radial overlap correction
which is updated to $\delta_{C2}$=\qty{0.310(14)}{\percent} \cite{townerPersonalCommunication2022} compared to the previous \qty{0.280(15)}{\percent} \cite{PhysRevC.102.045501}. The impact of this sizable change in $\delta_{C2}$ are summarised in Fig.~\ref{fig:Ft_shift}a and in Tab.~\ref{tab:summary_results}.

Despite $^{26m}$Al being the most accurately studied superallowed $\beta$ emitter, the corrected ${\mathcal{F}t}$ value is shifted by almost one full standard deviation to \qty{3071.4(10)}{\second}. Its high precision is maintained but, in terms of $R_c$ in the calculation of $\delta_C$, the value now stands on a solid experimental basis. 
The updated $\mathcal{F}t$ value of $^{26m}$Al also affects the $\overline{\mathcal{F}t}$ value, i.e. the weighted average over all 15 precisely studied superallowed $\beta$ emitters, which is shifted by one half of its statistical uncertainty
, see inset in Fig. 2a. To our knowledge, this represents the largest shift in the $\overline{\mathcal{F}t}$ value since 2009, see Fig.\ref{fig:Ft_shift}b.  This is a remarkable influence of a single experimental result on a quantity which is based on more than 200 individual measurements and which is dominated in its uncertainty by theoretical corrections.

Accounting for 0.57 s, this statistical uncertainty contains all experimental as well as those theoretical errors which scatter `randomly' from one superallowed transition to another. 
Previously, a single systematic theoretical uncertainty of  \qty{0.36}{\second} due to $\delta'_R$ had to be added affecting all superallowed $\beta$ emitters alike \cite{PhysRevC.91.025501}.
In these circumstances, the shift in the $\overline{\mathcal{F}t}$ value caused by the new charge radius of $^{26m}$Al would have corresponded to $\approx40\%$ of its total uncertainty. In the latest survey of superallowed $\beta$ decays \cite{PhysRevC.102.045501}, however, a systematic theoretical uncertainty of \qty{1.73}{\second} in $\delta_{NS}$ was newly introduced, reflecting uncertainties due to previously unaccounted contributions to the nuclear-structure dependent radiative corrections. 
This represents an almost three-fold increase of the theoretical error associated with $\delta_{NS}$ which now dominates the uncertainty in the $\overline{\mathcal{F}t}$ value. 
Considering our new charge radius of $^{26m}$Al, one thus obtains an $\overline{\mathcal{F}t}$ value of \qty{3071.96(185)}{\second}.

    \begin{table}[b]
    \caption{\label{tab:summary_results}
    Summary of the rms charge radius $R_c$, the radial overlap correction $\delta_{C2}$ and the $\mathcal{F}t$ value of $^{26m}$Al, the weighted average of the 15 superallowed $\beta$ emitters $\overline{\mathcal{F}t}$ and the result of the CKM unitarity test.}
    \begin{ruledtabular}
    \begin{tabular}{rrr}
    quantity & previous value & this work \\
    \hline
    $R_c$ & \qty{3.040(20)}{\femto\meter} \cite{PhysRevC.66.035501} & \qty[text-series-to-math]{3.130(15)}{\femto\meter}\\
    $\delta_{C2}$ & \qty{0.280(15)}{\percent} \cite{PhysRevC.102.045501} & \qty[text-series-to-math]{0.310(14)}{\percent}\\
    $\mathcal{F}t(^{26m}Al)$ & \qty{3072.4(11)}{\second} \cite{PhysRevC.102.045501} & \qty[text-series-to-math]{3071.4(10)}{\second}\\
    $\overline{\mathcal{F}t}$ & \qty{3072.24(185)}{\second} \cite{PhysRevC.102.045501} & \qty[text-series-to-math]{3071.96(185)}{\second}\\
    $\Delta_{CKM}$ & \num{152(70)e-5} \cite{PDG2022} & \num[text-series-to-math]{144(70)e-5}\\
    \end{tabular}
    \end{ruledtabular}
    \end{table}

The present work further implies 
a $\Delta_{CKM}$ in the unitarity test of the first row of the CKM matrix which is brought by
$\approx1/10~\sigma$ closer towards unitarity.
Although the magnitude of this change is too small to resolve the tension to CKM unitarity, it illustrates the importance of a comprehensive examination of all relevant ingredients to $V_{ud}$, especially theoretical corrections which involve nuclear-structure dependencies such as radiative and ISB corrections. In terms of 
$\delta_{C2}$,
there remain seven superallowed $\beta$ emitters
in which the nuclear charge radius is experimentally undetermined \cite{Fricke2004, Angeli2013}. Among those, $^{10}$C and $^{14}$O are of specific interest given their sensitivity to the Fierz interference term which relates to scalar contributions in $\beta$ decays. Moreover, it has recently been proposed to constrain models of ISB corrections by new, more precise measurements of charge radii in triplets of the isobaric analog  states, e.g. $^{38}$Ca - $^{38m}$K - $^{38}$Ar  \cite{SENG2023137654}. 

{\it Summary. ---}
    Collinear laser spectroscopy has been performed to determine the nuclear charge radius of $^{26m}$Al, the most precisely studied superallowed $\beta$ emitter. The obtained value  differs by 4.5 standard deviations from the extrapolation used in the calculation of the isospin-symmetry-breaking corrections \cite{PhysRevC.66.035501, PhysRevC.102.045501}.
    This notably impacts the corrected $\mathcal{F}t$ value in  $^{26m}$Al and, thus, the average of all $\mathcal{F}t$ values used in the extraction of V$_{ud}$.
    As demanded by the tension in CKM unitarity, this work contributes to the thorough examination of all nuclear-structure dependent corrections in superallowed $\beta$ decays. Stimulated by the present results, efforts to measure experimentally undetermined charge radii of other cases, for example $^{54}$Co at IGISOL/Jyv\"askyl\"a, are currently ongoing.

    We would like to express our gratitude to the ISOLDE collaboration and the ISOLDE technical teams, as well as the IGISOL collaboration and IGISOL technical teams for their support in the preparation and successful realisation of the experiments. We are thankful for all input and discussions that we received from Ian S. Towner to support this work. S.M-E. is grateful for fruitful discussions with G. Ball. 

    We acknowledge funding from the Federal Ministry of Education and Research under Contract No. 05P15RDCIA and 05P21RDCI1 and the Max-Planck Society, the Helmholtz International Center for FAIR (HICfor FAIR), and the EU Horizon 2020 research and innovation programme through ENSAR2 (Grant No. 654002), grant agreement no. 771036 (ERC CoG MAIDEN) and grant agreement no. 861198-LISA-H2020-MSCA-ITN-2019. We acknowledge the funding provided by the UK Science and Technology Facilities Council (STFC) Grants No. ST/P004598/1 and ST/L005794/1. This work was supported by the FWO Vlaanderen and KU Leuven project C14/22/104. TRIUMF receives federal funding via a contribution agreement with the National Research Council of Canada. A significant share of the research work described herein originates from R\&D carried out in the frame of the FAIR Phase-0 program of LASPEC/NUSTAR. 
    
\bibliography{references, referencesStephan}

\end{document}